\documentclass{aa}
\usepackage{graphics}
\usepackage{xspace}

\newcommand{\lgLrat}{\ensuremath{\lg{(L_{\rm x,qui}/L_{\rm bol})}}\xspace}

\begin{document}

\thesaurus{06(%
13.25.5; 
08.02.3, 08.09.2, 08.12.1, 08.01.2)} 

\title{An X-ray flare from the Lindroos binary system HD\,560}

\author{B. Stelzer\inst{1} 
\and N. Hu\'elamo\inst{1}}

\institute{Max-Planck-Institut f\"ur extraterrestrische Physik, 
 D-85740 Garching, Germany\\
}

\offprints{B. Stelzer}
\mail{stelzer@xray.mpe.mpg.de}
\date{Received $<$25 August 2000$>$ / Accepted $<$12 October 2000$>$}

\maketitle

\begin{abstract}

We report on a large X-ray flare from the Lindroos binary 
system HD\,560 during a pointed {\em ROSAT} PSPC observation. 
HD\,560 is composed of a B9 and a G5 type star. 
The late-type companions in bound Lindroos
binaries are Post T Tauri star (PTTS) candidates, i.e.
pre-main sequence stars on their final approach to the main sequence.
Strong (magnetic) activity is, therefore, expected for the G star, in contrast
to the late-B type component for which no mechanism of X-ray production is
known. The system is unresolved in the {\em ROSAT} image.

During outburst the count rate has changed by at least a factor of 2,
and presumably more (the maximum was not observed).
The peak luminosity measured is $\sim 10^{31}\,{\rm erg/s}$, comparable to 
the largest values derived for flares on pre-main sequence T Tauri stars. 
A two-temperature model for thermal emission from a hot, optically thin
plasma shows an increase of the higher temperature as a result of
coronal heating during the flare. The lower temperatures do not change
during outburst.


\end{abstract}

\section{Introduction}

T Tauri stars (TTS) are young, late-type stars contracting to the main sequence (MS). 
In their evolution pre-main sequence stars pass through 
a phase before reaching the MS,
defined as Post T Tauri stars (PTTS).
Herbig~(1978) first defined PTTS   
as young stars more evolved than Classical T Tauri stars
(CTTS) but still contracting to the MS.

PTTS are difficult to detect because they do not show extreme
properties which make them easy to identify: unlike CTTS, they do not
show IR or UV excesses and the H$\alpha$ line is not
longer seen in emission. Therefore, their identification relies
on the presence of the Li I (6708\AA) absorption line and the
Ca II chromospheric lines in their spectra as well as on
their X-ray emission.

Murphy (1969) was the first to propose that PTTS could be found in 
binary systems composed of early-type primaries
and late-type secondaries, given that the MS lifetime 
of high-mass stars is comparable to the contraction timescale
of late-type stars to the MS.  Gahm et al. (1983) and Lindroos (1985) 
carried out photometric and spectroscopic observations 
of 253 binary systems and Lindroos (1986) was able to isolate 
a sample of 78 visual binary systems with likely PTTS secondaries.

The X-ray emission from these PTTS candidates as observed by {\em ROSAT} 
has been studied by Schmitt et al. (1993) and Hu\'elamo et al. (2000). 
The derived X-ray luminosities  for
most of the late-type stars range between $\lg{L_{\rm x}} = 28 - 31\,[{\rm erg/s}]$.
These values are comparable to those reported 
for T Tauri stars stars (Neuh\"auser et al. 1995), showing that 
PTTS although somewhat more evolved are still strong X-ray emitters.
 
PMS late-type stars are also known to emit X-ray flares (see e.g. 
Feigelson \& DeCampli 1981, 
Montmerle et al. 1983, Stelzer et al. 2000).  Therefore, we have studied the 
X-ray lightcurves of all Lindroos PTTS candidates
observed by the {\em ROSAT} satellite 
(see Hu\'elamo et al. 2000 for
further details) in order to search for X-ray variability
due to activity episodes.

As a result of our study 
of the {\em ROSAT} PSPC and HRI lightcurves of the 
Lindroos systems, we have detected a large flare from HD\,560, a binary 
system composed of a B9 and a G5 star.  
A previous flaring episode from HD\,560 observed with {\em EXOSAT} 
was reported by Tagliaferri et al. (1988).
The {\em EXOSAT} Medium Energy experiment provided no spatial resolution,
however. Therefore, it was unclear whether to ascribe the event to HD\,560
or to the nearby Seyfert type I galaxy, QSO\,0007+107, 
located at $\alpha = 00^{\rm h}10^{\rm m}31.01^{\rm s}$ and 
$\delta = 10^\circ58^\prime 29.5^{\prime\prime}$. 
While Pounds \& Turner~(1987) attributed the flare to the galaxy, Tagliaferri
et al. (1988) argued in favor of HD\,560.
The observation of another flare with the {\em ROSAT} PSPC 
whose spatial resolution is sufficient to identify the flaring source with
HD\,560, supports the arguments put forth by Tagliaferri et al. (1988).
In this paper we provide an analysis of the {\em ROSAT} data of HD\,560,
and discuss the flare event.

\section{The {\em ROSAT} data}

The X-ray telescope and the instrumentation onboard
the {\em ROSAT} satellite are described in detail by Tr\"umper (1983),
Pfeffermann et al. (1988) and David et al. (1996).
The data presented in this work were obtained with the 
Position Sensitive Proportional Counter (PSPC) in pointed
mode. 

The spectral resolution of the PSPC (43\% 
at 0.93 keV) allows spectral analysis in three 
energy bands: \\
\noindent
- Soft: 0.1 to 0.4\,keV \\
- Hard 1: 0.5 to 0.9\,keV \\ 
- Hard 2: 0.9 to 2.0\,keV \\
We can obtain spectral information 
of our sources studying the X-ray hardness ratios ($HR$)
defined as follows:\\

\begin{equation} 
\; \; HR1=\frac{(H1+H2-S)}{(H1+H2+S)} \; \; \mbox{and} \;\;
      HR2=\frac{(H2-H1)}{(H2+H1)}
\end{equation}
where $H1$ and $H2$ are the counts observed in the Hard 1 
and Hard 2 bands, and $S$ are the counts observed in the
Soft band. Hence, $HR$ values can range from $-1$ to $+1$.

\section{The X-ray emission from HD\,560}

\begin{table*}
    \caption{PSPC Pointed Observations of HD\,560. See text for a
    description of the columns.}
    \label{tab:xray_info}
    \begin{flushleft}
     \begin{tabular}{l@{\hspace{3mm}}l@{\hspace{2mm}}l
      @{\hspace{2mm}}c@{\hspace{3mm}}r@{\hspace{3mm}}r@{\hspace{3mm}}r
      @{\hspace{3mm}}c@{\hspace{3mm}}c@{\hspace{3mm}}r} 
      \noalign{\smallskip}
      \hline
      \noalign{\smallskip}
      ROR  & \multicolumn{2}{c}{X-ray position} & \multicolumn{1}{c}{$\Delta$} & \multicolumn{1}{c}{Offaxis} & \multicolumn{1}{c}{Counts} & \multicolumn{1}{c}{Expo} & \multicolumn{1}{c}{HR1} & \multicolumn{1}{c}{HR2}  & \multicolumn{1}{c}{ML} \\
      number  & \multicolumn{1}{c}{$\alpha_{\rm 2000}$} & \multicolumn{1}{c}{$\delta_{\rm 2000}$} & \multicolumn{1}{c}{[\arcsec]} & \multicolumn{1}{c}{[\arcmin]} &  & \multicolumn{1}{c}{[sec]} &     &      &  \\    
      & \multicolumn{1}{c}{[h m s]} & \multicolumn{1}{c}{[\degr~\arcmin~\arcsec]} & & & & & & & \\     \hline
     \noalign{\smallskip}
      700503p & 00 10 02.48 & 11 08 41.2 & 5.6/6.8 & 12.32 & 1993.2$\pm$45.3 
      & 8238.3 & -0.04$\pm$0.02 & 0.05$\pm$0.03 & 7713.6 \\
      701092p & 00 10 02.47 & 11 08 40.4 & 6.1/6.0 & 12.31 & 4258.3$\pm$65.7 
      & 8121.8 & -0.08$\pm$0.02 & 0.12$\pm$0.02 & 18859.8 \\ \hline
\end{tabular}
\end{flushleft}
\end{table*}

\begin{figure} 
\begin{center}
 \resizebox{8.5cm}{!}{\includegraphics{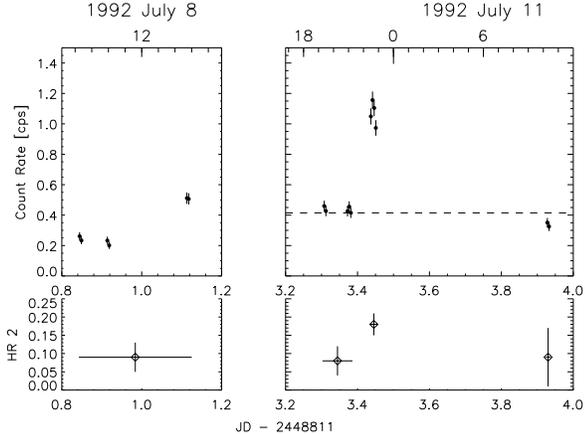}}
 \caption{ROR 701092: PSPC lightcurve and $HR2$ of the unresolved
 HD\,560 binary. Plotted are only fully exposed 400\,s bins.  
 The mean pre-flare count rate (dashed line) 
 was determined only from the 4th and 5th data intervals because 
 the quiescent level seems to be variable
 during the first part of the observation. For the estimation of the
 pre-flare level we have also made use of partially exposed time bins,
 which are not displayed in the figure. $HR2$ increases during the
 flare as a result of coronal heating.}
\label{fig:lc_hd560}
\end{center}
\end{figure}

HD\,560 was observed twice by the PSPC. The details of these
two observations are summarized in Table 1, where the {\em ROSAT}
observation request number (ROR)
is provided. Columns $2-4$ show the position of the
X-ray detection, its displacement with respect to the B9 and G5
components of HD\,560, respectively, and the distance of the 
X-ray source to the center of the detector. The latter number is important, 
given that the spatial resolution of the PSPC is off-axis dependent. 
The count rates, exposure times, and $HR$'s are given in 
columns~$5-8$. An estimation of the
probability of the detection according to the maximum likelihood (ML)
algorithm implemented in EXSAS (Extended Scientific Software Analysis
System, Zimmermann et al. 1995) 
is given in the last column. Note that $ML=5$ corresponds to a 2.7$\,\sigma$
signal over the background.

The spatial resolution of the PSPC ($\sim 25\arcsec$ at $1\,{\rm keV}$ 
at the center of the detector) does not allow to resolve the binary system 
($7.7^{\prime\prime}$ separation). Therefore, in both observations 
there is a single X-ray detection located at $\sim$6\,\arcsec~from the 
binary system (see Table~\ref{tab:xray_info}).

\section{The X-ray flare on HD\,560}

During the {\em ROSAT} PSPC observation ROR\,701092p the binary 
system HD\,560 showed a large flare.  The complete lightcurve
is shown in Fig. 1. Next to the large outburst (around JD
2448814.4) an increase of the count rate is observed from the 2nd
and 3rd data interval in Fig.~\ref{fig:lc_hd560}. Since the count rate
just before the eruption of the flare is about the same as that of the
3rd data interval, we suppose that between the 3rd and 5th data
interval we see the quiescent emission. However, due to poor data
sampling it is unknown for how long the emission indeed remained
stable at this level.  The strong variability besides the large 
flare might also reflect rotational modulation as a result of a spot
on the surface of one of the companions in the binary, or it could be some
kind of orbital effect.

With a binary separation of $7.7^{\prime\prime}$ the system is not resolved
in both PSPC observations.  
Late-B type stars are not expected to produce X-rays 
because they lack both convective envelope (necessary for solar-type dynamo
activity) and strong stellar winds (providing hot, shocked gas).
However, in earlier studies (Schmitt et~al. 1993, 
Hu\'elamo et~al. 2000)
X-rays were observed from both stars in those Lindroos pairs that were
resolved by the {\em ROSAT} HRI. While the typical 
$\lg{(L_{\rm x}/L_{\rm bol})}$
ratio for late-type stars is $-3$, early-type stars are 
characterized by $\lg{(L_{\rm x}/L_{\rm bol})} \sim -7$. 
Hu\'elamo et al. (2000) found that late-B type stars
show intermediate 
$L_{\rm x}/L_{\rm bol}$ values, but the hardness ratio is
similar to that of late-type stars, suggesting unresolved companions. 
In the case of HD\,560 we
derive $\lg{(L_{\rm x,qui}/L_{\rm bol})} = - 4.81$ using the
bolometric luminosity of the primary. Therefore it is likely that
the late-type component is the X-ray source, and, although we can not
exclude the opposite case,
we assume that the flare occurred on the G5 dwarf star.

We have used the count-to-energy conversion factor of 
$1.1\,10^{11}\,{\rm cts\,cm^2/erg}$ (Neuh\"auser et al. 1995) to compute
the X-ray luminosity. For the
{\em Hipparcos} distance of $100 \pm 9$\,pc the 
maximum count rate observed corresponds to a luminosity of $L_{\rm
x,max} = 1.3~10^{31}\,{\rm erg/s}$. 
The luminosity derived for the quiescent count rate is $L_{\rm x,qui} = 
4.5~10^{30}\,{\rm erg/s}$. 
We note, that an estimate of the possible 
contribution from the early-type primary 
to the quiescent emission is impossible.
Therefore, for the evaluation of the flare luminosity,
$L_f = L_{\rm max} - L_{\rm qui}$, we have studied two different 
cases: (a) all X-rays (during quiescence
and during the flare) come from the late-type  component, 
and (b) both stars in the binary system contribute the same level 
to the quiescent emission, but the additional emission 
during the flare is only due to the late-type star.
For these two cases we have found the following values for the flare
luminosity:

(a) $\; \; L_f = 8.1 \times 10^{30}\,{\rm erg/s}$  

(b) $\; \; L_f = 1.0 \times 10^{31}\,{\rm erg/s}$

No data is available for both rise and decay phase of the flare event.
We chose 400\,s for the binsize of the lightcurve displayed in
Fig.~\ref{fig:lc_hd560}. This is the best choice
for the bin time given the satellite orbit 
wobble, and in order to achieve good S/N in each bin 
and high time resolution. Furthermore, we find that choosing finer 
time resolution does not improve our knowledge about 
the temporal development of the
intensity. Due to the incomplete data sampling during the flare we 
can give only upper limits for both
rise and decay timescales: $\tau_{\rm rise} < 1.2\,{\rm h}$ and $\tau_{\rm decay} <
11.3\,{\rm h}$.

The time evolution of the PSPC hardness ratio $HR\,2$ is displayed
in the lower panel of Fig.~\ref{fig:lc_hd560}.  $HR\,2$ seems to
increase during the flare. However, the data are not conclusive in
this respect due to low statistics.  To verify whether the source
really hardened during the flare event, we have extracted X-ray
spectra for two time intervals representing the non-flaring state
(pre- and post- flare data) and the flare (data interval \#6).
Spectral models for thermal emission from a hot, optically thin plasma
(RS-model; Raymond \& Smith 1977) plus a photo-absorption term are
used to describe the spectra of both states.

\begin{table*}
\caption{Best fit parameters for the PSPC spectra of HD\,560 during
quiescence and during the flare. The column density of a photo-absorption
term was held fixed at $N_{\rm H} = 5.5~10^{19}\,{\rm cm^{-2}}$ in all
cases. In the 1-T model the abundance was left free. 
A good fit was obtained with 2-T models without varying the abundances.
The uncertainties represent the 90\,\% confidence level.}
\label{tab:specfits}
\begin{tabular}{lcccccccl} \hline
& Model & $T_1$        & $EM_1$           & Abundance & $T_2$       &
$EM_2$  & Abundance & $\chi^2_{\rm red}$ (dof) \\ 
&       & $[10^7\,{\rm K}]$ & $[10^{53}\,{\rm cm^{-3}}]$ &           &
$[10^7\,{\rm K}]$ & $[10^{53}\,{\rm cm^{-3}}]$ & & \\ \hline
\multicolumn{9}{l}{\bf Flare} \\
& 1-T   & $1.30 \pm 0.28$ & $12.8 \pm 1.1$ & $0.17 \pm 0.07$ &   -   &
-         & -    & 1.05 (50) \\
& 2-T   & $0.16 \pm 0.03$ & $1.0 \pm 0.3$ & $= 1.0$ & $1.47 \pm 0.20$ &
$5.2 \pm 1.1$ & $= 1.0$ & 1.21 (49) \\
\multicolumn{9}{l}{\bf Quiescence} \\
& 1-T   & $1.01 \pm 0.10$ & $4.8 \pm 0.3$ & $0.08 \pm 0.03$ &   -   &
-         & -    & 0.78 (64) \\
& 2-T   & $0.19 \pm 0.03$ & $0.5 \pm 0.1$ & $= 1.0$ & $1.23 \pm 0.08$ &
$1.2 \pm 0.2$ & $= 1.0$ & 0.93 (63) \\ \hline
\end{tabular}
\end{table*}

\begin{figure} 
\begin{center}
\resizebox{8.5cm}{!}{\includegraphics{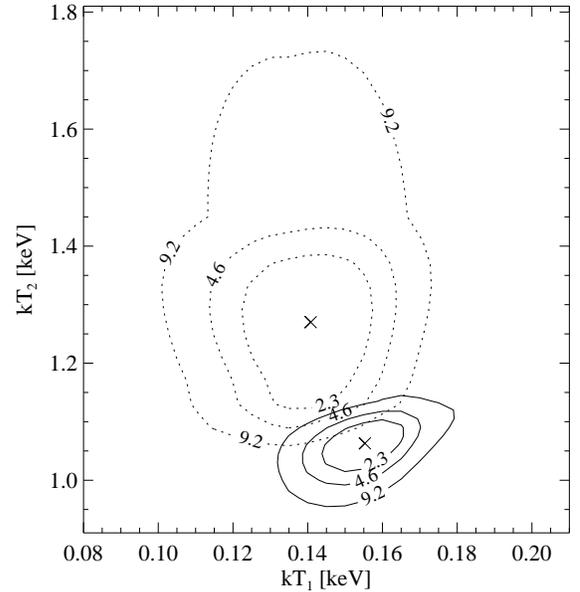}}
\caption{Contour plot comparing the quiescent and the flare state with
 respect to the two temperatures of the 2-T RS-model from
 Table~\ref{tab:specfits}. Three contours are drawn corresponding to $\Delta
 \chi^2 = 2.3$, $6.41$, and $9.21$. Solid lines for quiescence, dashed lines
 for the flare. Heating during the flare is expressed in an increase of the
 temperature in the hotter component. The cooler component remains little 
 affected by the outburst.}
\label{fig:contours}
\end{center}
\end{figure}

We obtain good fits for two-temperature (2-T) RS models with
abundances fixed to the solar value.  One-temperature (1-T) models
with solar abundance do not lead to acceptable fits ($\chi^2_{\rm red}
> 1.6$ for both flare and quiescent data).  The F-test reveals that
the significance of the improvement of the fit when the second
temperature component is introduced is higher than 0.99.  Only if the
abundance is allowed to vary during the fitting process the quality of
the fit is comparable to that of the 2-T models with fixed abundance.
In all cases the column density $N_{\rm H}$ was held fixed at
$5.5~10^{19}\,{\rm cm^{-2}}$ (Bergh\"ofer et al. 1996).  In
Table~\ref{tab:specfits} the best fit parameters from the 1-T model
with free abundance are compared to the result of the 2-T model with
fixed abundance for both quiescent and flare spectra.  
In the 1-T model only a marginal temperature increase during the flare is
observed despite the increase in spectral
hardness (see Fig.~\ref{fig:lc_hd560}).
The contour
plot for the two temperatures of the 2-T model is shown in
Fig.~\ref{fig:contours}. While the uncertainties make flare and
quiescent state undistinguishable concerning $kT$ of the cooler
component, the higher of the two temperatures is clearly representing
the plasma heated during the flare.

\section{Results and conclusions}

We have presented a large X-ray flare from the Lindroos binary system HD\,560
observed with the {\em ROSAT} PSPC. 
Earlier observations of a flare by {\em EXOSAT} had been reported, but left
the identification with HD\,560 ambiguous due to the lack of spatial
resolution of the Medium Energy experiment.

The maximum X-ray luminosity during the {\em ROSAT} flare corresponds to 
$\sim 10^{31}$\,erg/s, a value typical for young late-type stars. 
We have computed the bolometric luminosity for the G and the B star in
HD\,560 from the V magnitudes (Lindroos 1985)
and the bolometric correction listed by Schmidt-Kaler (1982).
Using the observed quiescent emission we derive the ratio of X-ray to
bolometric luminosity \lgLrat
for primary and secondary: $(\lgLrat)_{\rm prim} = -4.81$ and
$(\lgLrat)_{\rm sec} = -2.76$. The latter value agrees well with the
`canonical' value of $-3$ for late-type stars, while the value for the
primary is neither typical for late-type nor for early-type stars.
Therefore, supposedly, the flare has erupted on the G5 component of the
system. However, the binary is unresolved, and it can not be ruled out
that the flare originated on the B9 star. Whether or not late-B type stars
emit X-rays is an unsolved problem.

We have analysed X-ray spectra for 
both the quiescent and the flare phase, and find acceptable fits for both
phases with a two-temperature Raymond-Smith model.
During the flare the higher temperature showed an increase with 
respect to the quiescent stage, while the lower temperature remained
unaffected. 
A one-temperature 
model with free abundance provides fits of comparable quality.
However, the rather similar temperature derived for the
flare and the quiescent state with the one-temperature model 
is difficult to reconcile with 
coronal heating expected during flare outburst.

\begin{acknowledgements}

The archived {\em ROSAT} observations investigated here
were performed by R. Staubert. The {\em ROSAT} project is supported by the 
Max-Planck-Society and the German Government (DLR/BMBF).

\end{acknowledgements}

\end{document}